\documentclass[amsmath,amssymb,showpacs,showkeywords,twocolumn]{revtex4}
\usepackage{graphicx}
\usepackage{times}
\usepackage{mathrsfs}
\usepackage{amsmath}
\usepackage{graphicx}
\usepackage{epsfig}
\usepackage{dcolumn}
\usepackage{bm}
\usepackage{multirow}
\usepackage{pdfpages}
\usepackage[utf8x]{inputenc}
\DeclareUnicodeCharacter{2212}{-}
\begin{document}
\title{Quantum Capacitance and Electronic Properties of a Hexagonal Boron Nitride based FET Gas Sensor}

\author{Saumen Acharjee\footnote{saumenacharjee@dibru.ac.in}}
\affiliation{Department of Physics, Dibrugarh University, Dibrugarh 786 004, 
Assam, India}

\begin{abstract}
We present a comprehensive theoretical investigation of gas sensing in monolayer hexagonal boron nitride (h-BN) based field-effect transistors (FET) using the non-equilibrium Green function formalism and Landauer–B\"{u}ttiker approach. Moving beyond conventional density functional theory analyses, our framework captures the full device level response by incorporating field-dependent quantum transport and temperature effects. We model the impact of NO, H$_2$S, HF and CO$_2$ gases on the band structure and density of states (DOS), carrier concentration, quantum capacitance and I-V characteristics. The results indicate that CO$_2$ followed by NO induce strongest perturbations via mid-gap states and band edge shifts, leading to the appearance of asymmetric Van-Hove singularities with enhanced carrier modulation and quantum capacitance. It is observed that HF induce moderate perturbation while H$_2$S induce weakest response for all temperature and biasing condition. It is found that an applied vertical electric field narrows the band gap via the Stark effect, further boosting mobility and tunability. Temperature influences sensing response by enhancing charge transfer at moderate levels and causing desorption at higher temperatures.  We found that CO$_2$ consistently show the highest sensitivity and selectivity followed by NO and HF, while H$_2$S display the weakest response. This study offers a comprehensive framework to engineer h-BN based FET sensors by harnessing intrinsic band modulation and quantum capacitance for molecule discrimination and temperature optimization.
\end{abstract}

\maketitle

\section{Introduction}

Over the past two decades, two-dimensional (2D) materials have emerged as a transformative platform for next-generation electronics, energy devices and sensors due to their exceptional structural, electronic and surface properties \cite{novoselove,lemme,liu,xu,quellmalz,illarionov,fuechsle,acharjee, acharjee2}. The electronic \cite{balandin,lee,zhang}, mechanical \cite{pop}, thermal \cite{sang,pop} and optical properties \cite{pop} are dramatically modified due to the broken translational symmetry inherent in 2D monolayers, leading to edge-localized and surface-sensitive electronic states \cite{dutta,borca,siao}.  Such atomic-scale phenomena have stimulated intense interest in low-dimensional systems for functional device applications. In particular, the development of advanced microfabrication and exfoliation techniques have allowed for the extraction of atomically thin layers from bulk crystals, paving the way for the emergence of 2D materials as a major research frontier \cite{novoselov2,butler}. A landmark was the discovery of graphene, an individual layer of carbon atoms in a honeycomb lattice. Graphene exhibits exceptional carrier mobility, thermal conductivity and mechanical strength with all physical phenomena confined in one atomic layer \cite{geim, morozov,engels}. These exceptional properties are due to its van der Waals nature and the consequent quantum confinement, together providing novel control over the band structure and transport properties \cite{geim2}.

While early research largely centered on graphene, its intrinsic gapless band structure limits its application in electronic switching and selective sensing \cite{geim}. In contrast, hexagonal boron nitride (h-BN), an insulating counterpart to graphene offers a wide, direct bandgap ($\sim$ 5.9~eV) \cite{museur,verma,watanabe,cassabois},  excellent thermal \cite{sichel} and mechanical stability \cite{song}, thereby making it highly suitable for high-temperature and chemically harsh sensing environments \cite{watanabe,sichel,cassabois}. 
Moreover, the partially ionic nature of the B–N bond introduces local surface polarization, enhancing its interaction with polar gas molecules \cite{zuo}. Additionally, the wide
band gap allows for strong modulation of electronic properties upon adsorption of specific gases, particularly those
involving charge transfer. Recent computational and experimental studies have confirmed that monolayer and few layer h‑BN exhibit selective adsorption and strong interaction with toxic gases such as NO$_2$,  CO$_2$ and NO \cite{phung,kim2013,kim2024}. It is driven primarily by electrostatic and orbital level alignment mechanisms \cite{sharma,cai,kalwar, rahimi}. 

Apart from that, unlike many transition metal dichalcogenides (TMDs) that suffer from structural defects, instability under ambient or high effective mass conditions \cite{wang}, h-BN maintains atomically smooth surfaces and has low trap densities ensuring high-quality interfaces in electronic devices \cite{dean,novoselov2016}. Moreover, its insulating nature allows it to serve not only as an active sensing layer but also as an ultra-thin dielectric substrate or capping layer in heterostructures \cite{liu2014, britnell}. Compared to TMDs like MoS$_2$ or WS$_2$, which are typically semiconducting with narrower bandgaps, h-BN exhibits superior robustness, making it ideal for applications requiring long-term environmental stability, minimal leakage currents and compatibility with complementary 2D materials in van der Waals heterostructures \cite{geim2}. These attributes position h-BN as a promising alternative to graphene and other 2D materials in advanced sensing, nanoelectronic and optoelectronic platforms. Although there has been significant research on gas sensing using h-BN, most studies focus on the electronic structure and adsorption properties using density functional theory (DFT) \cite{phung,kim2013,kim2024,sharma,rahimi,kalwar}. 
These studies are invaluable for understanding gas material interactions at the atomic scale; however, they do not address how such interactions influence real world device behavior of h-BN-based sensors. In particular, how gas adsorption influences the actual current flow, quantum capacitance and carrier concentration in a device under typical operating conditions are yet to be explored. Unlike equilibrium DFT approaches, the nonequilibrium Green function (NEGF) formalism \cite{baym,keldysh,dutta2005} accounts for open boundary conditions and non-equilibrium carrier dynamics, enabling direct simulation of how gas adsorption modifies current, carrier injection and gate control under real operating conditions \cite{ozdemir, kanrar, alaee, raval,le}. To address this gap, our work develops a quantum transport model based on NEGF to directly link molecular adsorption to observable changes in current, carrier density and quantum capacitance in h-BN field-effect transistors (FETs) under bias and gate fields \cite{kanrar,raval}. By combining molecular-level insights with realistic device-scale simulations, we can predict how adsorbed gases impact the performance of h-BN sensors, providing a more comprehensive and accurate framework for designing and optimizing these devices.

The organization of this paper is as follows: In section II, we present the our model and formalism based on a gas-functionalized tight-binding Hamiltonian for monolayer h-BN, followed by implementation of the NEGF formalism to model quantum transport through the FET-based gas sensing device. Section III-VI, discusses the numerical results for the electronic density of states, carrier concentration, quantum capacitance and current–voltage (I-V) characteristics, highlighting the effects of gas adsorption on the device performance. Finally, section VII summarizes our key findings and outlines potential directions for experimental validation and future research on h-BN–based nanoscale gas sensors.

\begin{figure}[t]
\centerline
\centerline{
\includegraphics[scale=0.28]{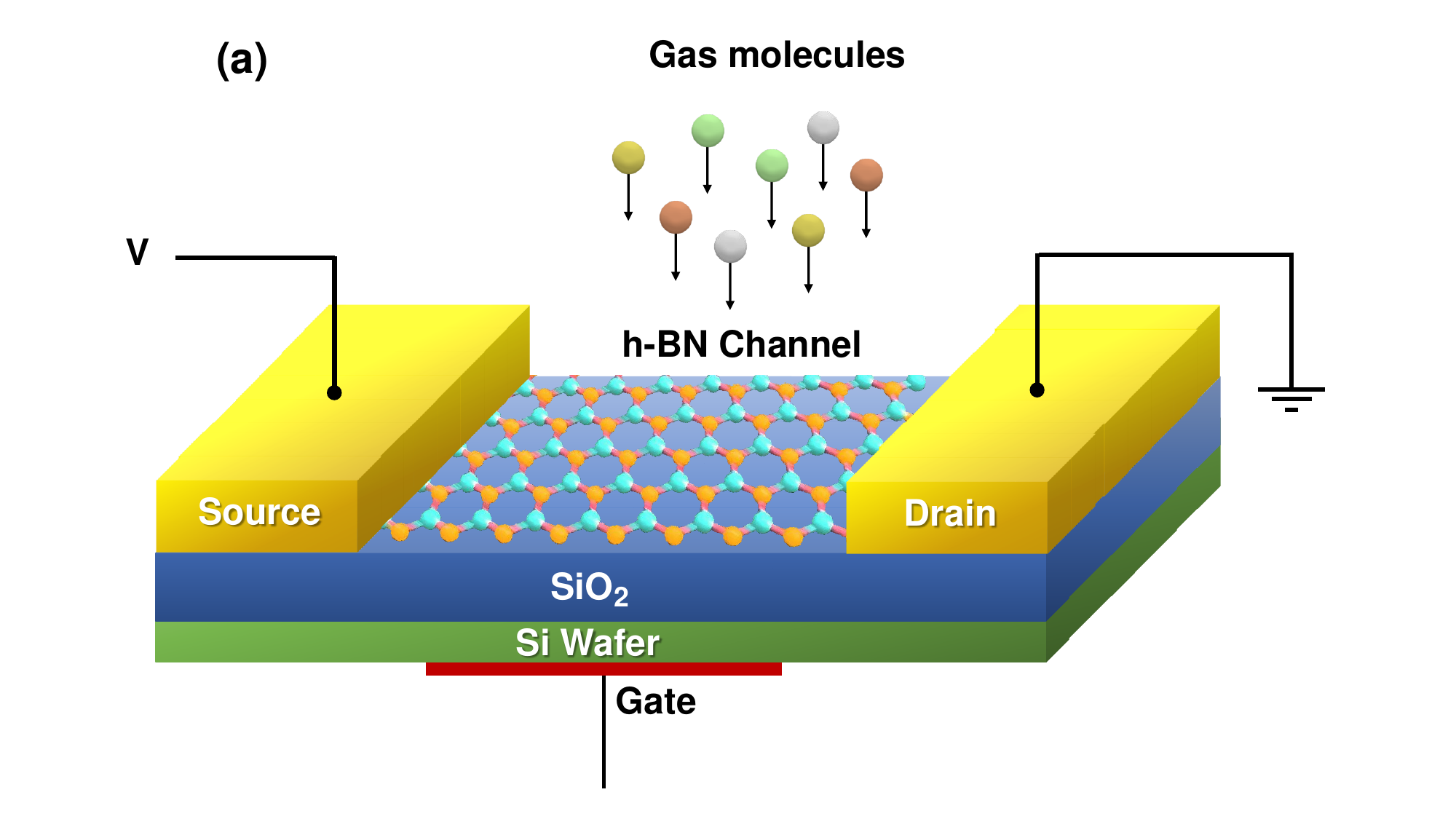}
\includegraphics[scale=0.28]{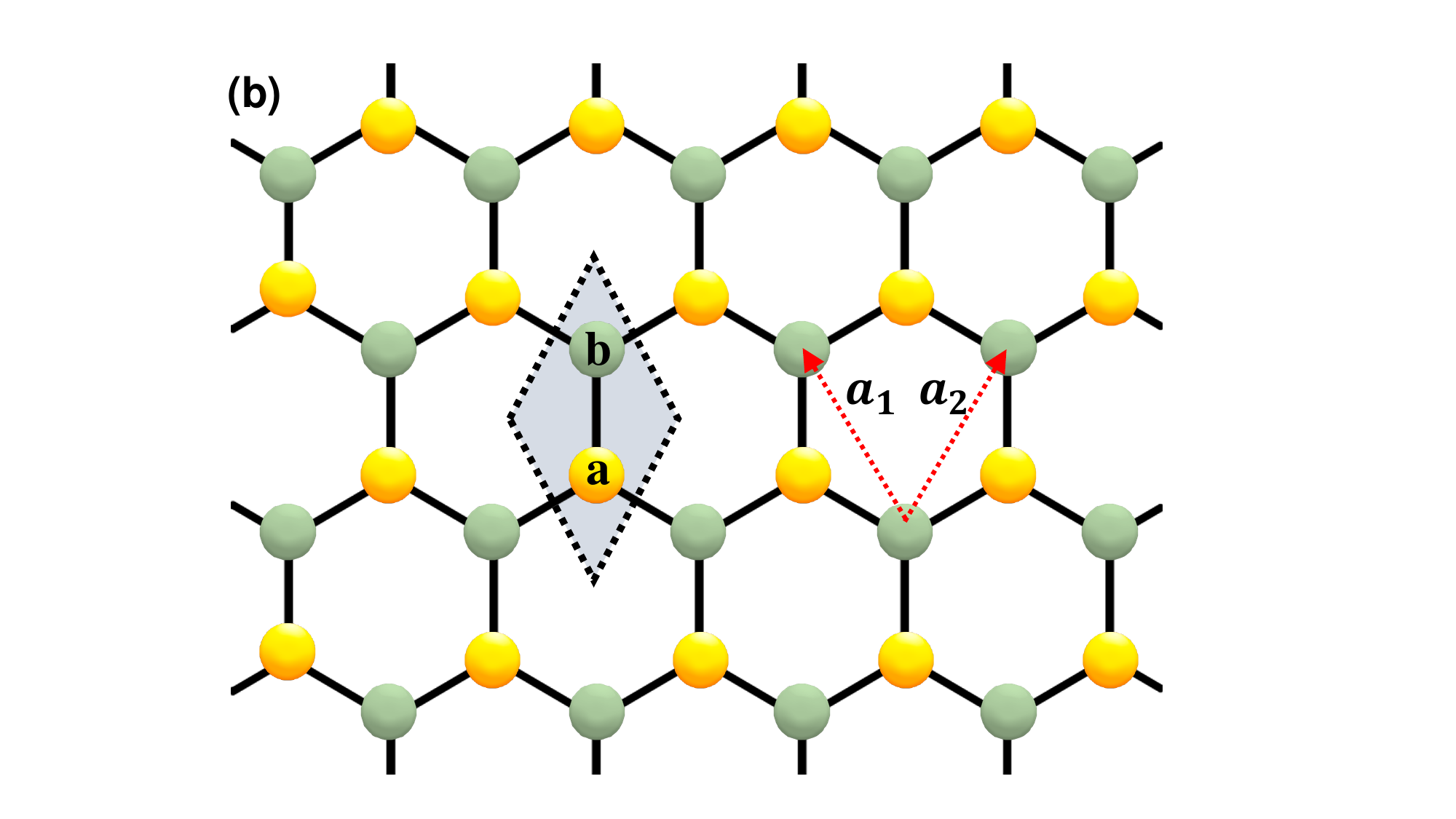}
}
\caption{(a) Schematic of a h-BN Field Effect Transistor (FET) for gas sensing device. The device is connected to two semi-infinite leads (left and right) and is subjected to an external bias voltage $V$. (b) Top view of the h-BN geometry. The h-BN unit cell (shaded blue area) is made up of one boron (yellow) atom and its nearest neighbor nitrogen (green) atom.
of the honeycomb structure, labeled as $a$ and $b$ respectively.}
\label{fig1} 
\end{figure}

\section{Model and Formalism}

We consider a monolayer h-BN FET based gas sensor as presented schematically in Fig. \ref{fig1}. In this setup, the h-BN monolayer serves as the active channel, interfaced with source and drain electrodes on a SiO$_2$/Si back-gated substrate.  When target gas molecules are exposed, the local electronic structure is perturbed due to the adsorption of the gases which in turn modulates its carrier transport properties. This change enables electrostatic detection through gate-tunable conductance.
The atomic arrangement of h-BN consists of alternating boron and nitrogen atoms arranged in a planar honeycomb lattice generated by the lattice vectors $\vec{a}_1 = \left\{\frac{3 a_0}{2},\frac{\sqrt{3} a_0}{2}\right\}$ and $\vec{a}_2 = \left\{\frac{3 a_0}{2}, -\frac{\sqrt{3} a_0}{2}\right\}$ with the lattice constant $a_0 = 2.5$ \AA  and B-N bond length is approximately 1.45 \AA as shown in Fig. \ref{fig1}(b). The atomically thin character of h-BN offers high surface sensitivity, making it an ideal platform for charge-based sensing. In contrast to graphene, h-BN exhibits a large intrinsic band gap arising from the ionic character of the B-N bond, which breaks sublattice symmetry and thereby suppresses $\pi$-conjugation. The electronic states near the band edges are dominated by $ p_z$ orbitals. Furthermore, the lack of band dispersion leads to strong localization effects under external perturbations.

We investigate quantum transport through a nanoscale device connected to semi-infinite leads under a finite drain source voltage $V_{ds}$, employing the NEGF formalism. To capture the effect of gas adsorption, we consider a tight-binding model on the h-BN lattice that captures both local orbital characteristics and hopping integrals. The total Hamiltonian of the system can be written as
\begin{equation}
    \mathcal{H} = \mathcal{H}_\text{D} + \mathcal{H}_\text{L} + \mathcal{H}_\text{R} + \mathcal{H}_\text{LD} + \mathcal{H}_\text{DR}
    \label{eq1}
\end{equation}
where,  $\mathcal{H}_\text{D}$ is the device Hamiltonian, $\mathcal{H}_\text{L}$ and $\mathcal{H}_\text{R}$ describe the left and right leads respectively and $\mathcal{H}_\text{LD}$ ($\mathcal{H}_\text{DR}$) represents the coupling between the device and the left (right) lead.

The tight-binding Hamiltonian of the h-BN system in the device region is given by \cite{dutta2005}
\begin{equation}
    \mathcal{H}_\text{D} = \sum_i \epsilon_i c_i^\dagger c_i + \sum_{i\neq j} t_{ij} c_i^\dagger c_j
    \label{eq2}
\end{equation}
where, $\epsilon_i$ is the on-site energy and $t_{ij}$ are the hopping energy between atom $i$ and atom $j$. 

The matrix form of the Schr\"{o}dinger equation can be expressed as an eigenvalue problem:
\begin{equation}
    \sum_n c_n \langle m | \mathcal{H} | n \rangle = \mathcal{E} \sum_n c_n \langle m | n \rangle
    \label{eq3}
\end{equation}
where, $\mathcal{E}$ are the energy eigenvalues and $c_n$ are the coefficients of the linear combination of basis states in unit cell $n$. Using the orthogonality condition $\langle m | n \rangle = \delta_{mn}$, the equation simplifies to
\begin{equation}
    \sum_n \mathcal{H}_{mn} c_n = \mathcal{E} c_m
    \label{eq4}
\end{equation}

The matrix element between neighboring atoms appears as upper or lower diagonal terms. Assuming the hopping energy to be $-t$, we can write:
\begin{equation}
    \mathcal{H}_{mm} c_m = \mathcal{E} c_m
     \label{eq5}
\end{equation}

To understand the molecular adsorption effects, we consider a monolayer h-BN sheet with $N$ lattice sites. Each h-BN unit cell has four neighboring unit cells. According to the tight-binding model the wavefunction for the $n$-th unit cell can be written as $\psi_n = \psi_0 e^{i \vec{k} \cdot \vec{d}_n}$, where, $\vec{k}$ is the wave vector and $\vec{d}_n$ is the position vector of the $n^{th}$ unit cell. The Hamiltonian matrix for h-BN, with two atoms and one adsorbed molecule per unit cell is of the dimension $3 \times 3$ and the matrices describing interactions with neighboring cells can be written as 
\begin{align}
    \mathcal{H}_{n,1} &= 
    \begin{bmatrix}
        0 & t & 0 \\
        0 & 0 & 0 \\
        0 & 0 & 0
    \end{bmatrix}, \quad
    \mathcal{H}_{n,2} = 
    \begin{bmatrix}
        0 & 0 & 0 \\
        t & 0 & 0 \\
        0 & 0 & 0
    \end{bmatrix} 
    \label{eq6}\\
    \mathcal{H}_{n,3} &= 
    \begin{bmatrix}
        0 & 0 & 0 \\
        t & 0 & 0 \\
        0 & 0 & 0
    \end{bmatrix}, \quad
    \mathcal{H}_{n,4} = 
    \begin{bmatrix}
        0 & t & 0 \\
        0 & 0 & 0 \\
        0 & 0 & 0
    \end{bmatrix}
    \label{eq7}
\end{align}

The on-site Hamiltonian of unit cell $n$ including gas molecule adsorption is:
\begin{equation}
    \mathcal{H}_{n,n} = 
    \begin{bmatrix}
        E_0 & t & 0 \\
        t & E_1 & t' \\
        0 & t' & E_0'
    \end{bmatrix}
    \label{eq8}
\end{equation}
where, $t$ is the hopping between Boron and Nitrogen atoms, $t'$ is the hopping between Boron and the adsorbed molecule. Here $E_0$, $E_1$ and $E_0'$ are the on-site energies of nitrogen, boron and the adsorbed molecules respectively. The on-site energies $E_0$ and $E_1$ from homogeneous model are -2.55 eV and 2.46 eV respectively \cite{kalwar}. Gas adsorption modifies the electronic structure of h-BN by orbital hybridization primarily with Boron atoms due to their lower electronegativity. This is modeled by an added site per unit cell, with hopping $t'$ describing coupling between adsorbed molecule with Boron atom. The hybridization strength depends on orbital overlap, which varies with adsorption distance. In addition, charge transfer and dipole formation further shift on-site energies, altering the potential landscape. These interactions breaks local symmetry and thus modulate the band gap \cite{rahimi}. 

Due to lattice periodicity, the Hamiltonian in momentum space can be obtained by using the relation \cite{dutta2005,acharjee2}
\begin{equation}
h(\vec{k}) = \sum_{m = 1}^n \mathcal{H}_{n,m} e^{i \vec{k} \cdot (\vec{d}_n - \vec{d}_m)}
\label{eq9}
\end{equation}

Using of Eqs. (\ref{eq5}) - (\ref{eq8}) in Eq. (\ref{eq4}),  we can write 
\begin{align}
    h(\vec{k}) &= 
    \begin{bmatrix}
        E_0  &  t f(k) & 0 \\
        t f^\ast (k) & E_1  & t' \\
        0 & t' & E_0' 
    \end{bmatrix} 
    \label{eq10}
\end{align}
where, $f(\vec{k}) = 1 +  e^{-i k a_1} +  e^{-i k a}$. For non trivial solutions, $ \det[h(\vec{k}) - \mathcal{E} \hat{I}] = 0$, which yields three energy bands for each $\vec{k}$, can be obtained as follows
\begin{equation}
(E_0 - \mathcal{E})\left\{(E_1 - \mathcal{E})(E_0' - \mathcal{E}) 
-  t'^2\right\} 
- t^2 |f(\vec{k})|^2 (E_0' - \mathcal{E}) = 0
\label{eq11}
\end{equation}
where, $\hat{I}$ is a $3\times 3$ identity matrix and $\mathcal{E}$ are the energy eigenvalues, can be obtained by solving the Eq. (\ref{eq11}). It is to be noted that the widening of the band gap is dependent on the hopping energy of a boron atom with the adsorbed gas molecule. The hopping energies can be calculated using the relation \cite{acharjee2}:
\begin{equation}
    t_\text{xy} = t \left( \frac{a_0}{d_\text{xy}} \right)
    \label{eq12}
\end{equation}
where, $t_\text{xy}$ is the modified hopping energy, $t$ is the pristine hopping energy parameter, $a_0$ is the lattice constant of h-BN, and $d_\text{xy}$ is the distance between the h-BN surface and the adsorbed gas molecule. The hopping
parameters for different gases are given in Table \ref{tab1}.

\subsection{Surface Green's Function of the Semi-Infinite Leads}

Considering the each lead as a semi-infinite periodic array of identical unit cells, the Hamiltonian of the left (right) lead in Eq. (\ref{eq1}) is given by \cite{dutta2005}
\begin{equation}
    \mathcal{H}_{\mathrm{L(R)}} = \sum_n c_n^\dagger \mathcal{H}_{n,n} c_n + \left( c_n^\dagger \mathcal{H}_{n,1} c_{n+1} + \mathrm{H.c.} \right)
    \label{eq13}
\end{equation}
where, $\mathcal{H}_{n,n}$ and $\mathcal{H}_{n,1}$ are the intra-cell and inter-cell Hamiltonians respectively as defined in Eqs. (\ref{eq6}) - (\ref{eq8}) and $c_n$ is the annihilation operator for the $n^\text{th}$ unit cell. To understand the influence of the semi-infinite left and right leads on the device region, we computed the surface Green's function of the leads. 

Considering the periodicity of the system in spatial direction we transform it to momentum space by performing a Fourier transformation. The retarded surface Green's function of a semi-infinite lead projected onto the surface unit cell in momentum space can be defined as
\begin{equation}
    \mathcal{G}_\text{surf}(\vec{k}, \mathcal{E}) = \langle 0 | \left[ (\mathcal{E} + i\delta)\hat{I} - \mathcal{H}_\text{L(R)}(\vec{k}) \right]^{-1} | 0 \rangle
    \label{eq14}
\end{equation}
where,  $\mathcal{H}_\text{L(R)}(\vec{k})$  is the momentum-resolved Hamiltonian of the semi-infinite left (right) lead and  $|0\rangle$  denotes the basis states in the surface unit cell.

Since direct inversion of infinite-dimensional matrices is numerically prohibitive, we compute the surface Green's function $\mathcal{G}_\text{surf}(\vec{k}, \mathcal{E})$ using the iterative scheme developed by Sancho et al., \cite{sancho1,sancho2}, which is a recursive Green’s function method based on Dyson’s equation. This renormalization-based approach allows efficient and stable calculation of semi-infinite lead self-energies by systematically decimating degrees of freedom in the principal layer representation.

\subsection{Self-Energy Evaluation }

With the computed surface Green’s function in  $\vec{k}$-space, the self-energies due to the left and right leads are given by
\begin{equation}
    \Sigma_\text{L(R)}(\vec{k}, \mathcal{E}) = \mathcal{H}_\text{LD(DR)}^\dagger(\vec{k}) \, \mathcal{G}_\text{surf}(\vec{k}, \mathcal{E}) \, \mathcal{H}_\text{LD(DR)}(\vec{k})
    \label{eq15}
\end{equation}
where, $\mathcal{H}_\text{LD(DR)}(\vec{k})$  represents the  $\vec{k}$-dependent coupling between the device and the left (right) lead.

Thus the retarded Green's function of the device incorporating the effects of both leads can be written as \cite{keldysh,dutta2005}
\begin{equation}
    \mathcal{G}(\vec{k}, \mathcal{E}) = \left[ (\mathcal{E} + i\delta)\hat{I} - \mathcal{H}_\text{D}(\vec{k}) - \Sigma_\text{L}(\vec{k}, \mathcal{E}) - \Sigma_\text{R}(\vec{k}, \mathcal{E}) \right]^{-1}
    \label{eq16}
\end{equation}
where $\delta$ is the broadening factor.

To obtain physical observables such as the density of states or transmission function, we perform a summation over Brillouin zone $\vec{k}$.
\begin{equation}
    \mathcal{G}(\mathcal{E}) = \frac{1}{N_k} \sum_{\vec{k}} \mathcal{G}(\vec{k}, \mathcal{E})
    \label{eq17}
\end{equation}
where, $N_k$ denotes the total number of $\vec{k}$-points sampled in the Brillouin zone.

\begin{table}[t]
\caption{\label{tab1}
Adsorbate distance from h-BN sheet and the corresponding modified hopping energy for different gas molecules \cite{kalwar}.}
\begin{ruledtabular}
\begin{tabular}{lcc}
Adsorption Type & \begin{tabular}[c]{@{}c@{}}Adsorbate Distance\\ from h-BN (\AA)\end{tabular} & Hopping Energy (eV) \\
\hline
CO$_2$ & $-2.06$   & $0.16$ \\
H$_2$S & $-2.6785$ & $0.34$ \\
NO     & $-2.1126$ & $0.19$ \\
HF     & $-2.39$   & $0.27$ \\
\end{tabular}
\end{ruledtabular}
\end{table}

\begin{figure}[hbt]
\centerline
\centerline{ 
\includegraphics[scale=0.23]{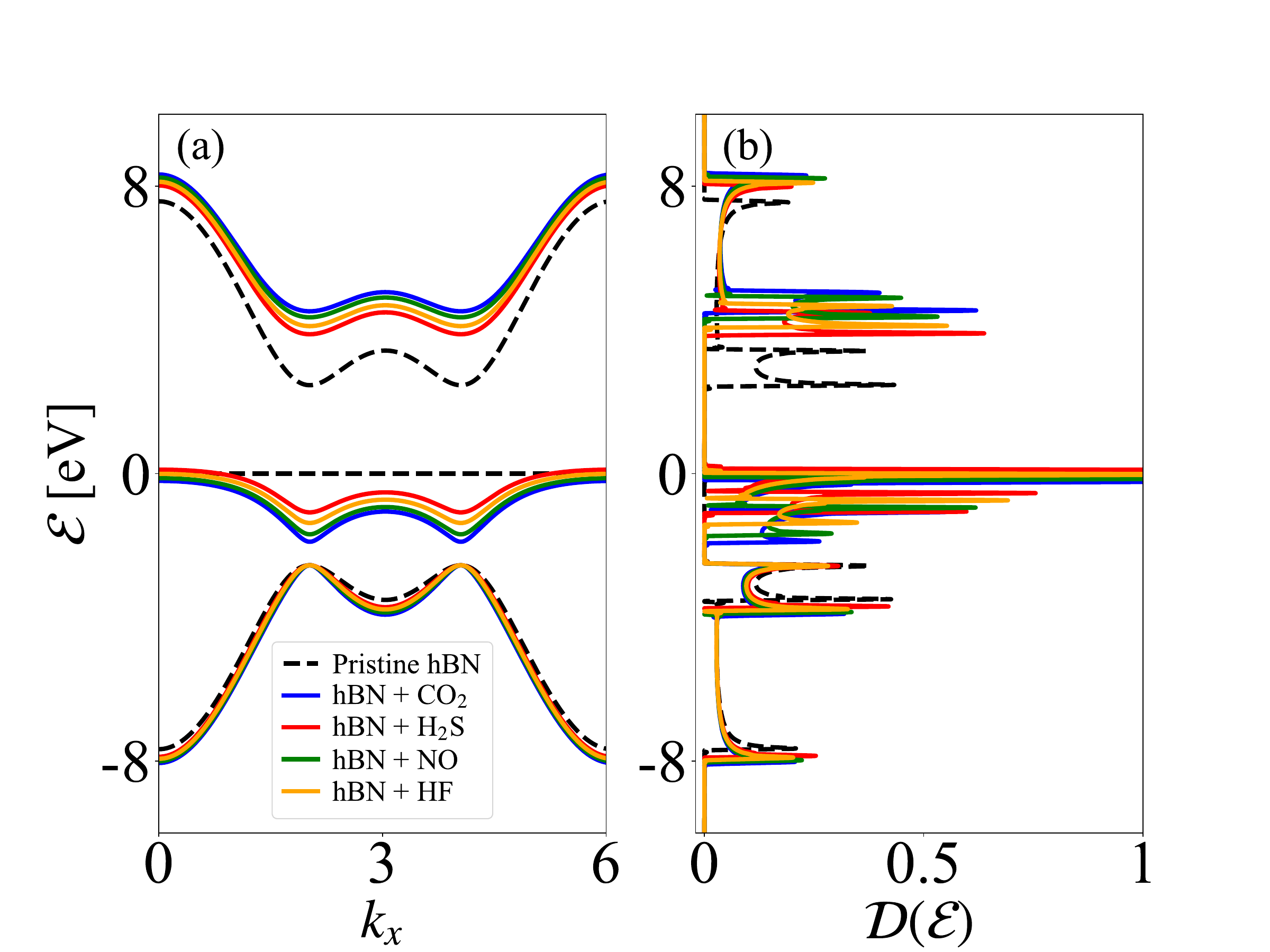}
}
\caption{(a) Band structure of Pristine h-BN and in presence of gases in absence of external biasing (b) The corresponding density of states of the h-BN system.}
\label{fig2}
\end{figure}

\section{Density of States}
Using the Green function, Eq. \ref{eq16}, we can calculate Density of States (DOS) for both pristine h-BN and in the presence of gases. The total DOS is obtained by summing over the full Brillouin zone. The DOS of the h-BN based FET device is given by \cite{le,dutta2005}
\begin{equation}
    \mathcal{D}(\mathcal{E}) = -\frac{1}{N_c \pi} \sum_{\vec{k}} \text{Im} \left[ \text{Tr} \, \mathcal{G}(\vec{k}, \mathcal{E}) \right]
\label{eq18}
\end{equation}
where, $N_c$ is the number of unit cells.

After simplification, the elements of the Green function matrix $\mathcal{G}_0$ in the device region can be written as
\begin{equation}
    \mathcal{G}_0(\vec{k}, \mathcal{E}) =
    \begin{bmatrix}
        \mathcal{G}_{11}^0 & \mathcal{G}_{12}^0 & \mathcal{G}_{13}^0 \\
        \mathcal{G}_{21}^0 & \mathcal{G}_{22}^0 & \mathcal{G}_{23}^0 \\
        \mathcal{G}_{31}^0 & \mathcal{G}_{32}^0 & \mathcal{G}_{33}^0 \\
    \end{bmatrix}
    \label{eq19}
\end{equation}
and the corresponding DOS can be obtained as follows:
\begin{equation}
    \mathcal{D}_0(\mathcal{E}) = -\frac{1}{N_c \pi} \sum_{\vec{k}} \text{Im} \left[ \mathcal{G}_{11}^0 + \mathcal{G}_{22}^0 + \mathcal{G}_{33}^0 \right]
    \label{eq20}
\end{equation}

Representative expressions for elements:
\begin{align}
    \mathcal{G}_{11}^0 &= \frac{E_0'E_1 - t'^2 - E_0'\epsilon + E_1\epsilon + \epsilon^2}{(E_0  - \epsilon  ) (t'^2+ \mathcal{P})} \\
    \mathcal{G}_{22}^0 &= \frac{E_0E_1 - E_0\epsilon + E_0'\epsilon + \epsilon^2}{(E_0  - \epsilon  ) (t'^2+ \mathcal{P})} \\
    \mathcal{G}_{33}^0 &= \frac{\mathcal{P}}{(E_0  - \epsilon  ) (t'^2+ \mathcal{P})}
\end{align}
where, we define $\mathcal{P}$ and $\mathcal{Q}$ as:
\[\begin{aligned}
\mathcal{P} &= E_0 E_1 -\epsilon (E_0  - E_1) + \epsilon^2  - t^2 +\mathcal{Q} \\
    \mathcal{Q} &= - 2t^2- 2t^2 \left\{\cos\left( \frac{k a_1}{2} \right) + \cos\left( \frac{k a_2}{2} \right)\right\}  \\
&\quad- 4t^2 \cos\left( \frac{k a_1}{2} \right) \cos\left( \frac{k a_2}{2}\right)
\end{aligned}\]

\begin{figure*}[t]
\centering
\includegraphics[scale=0.23]{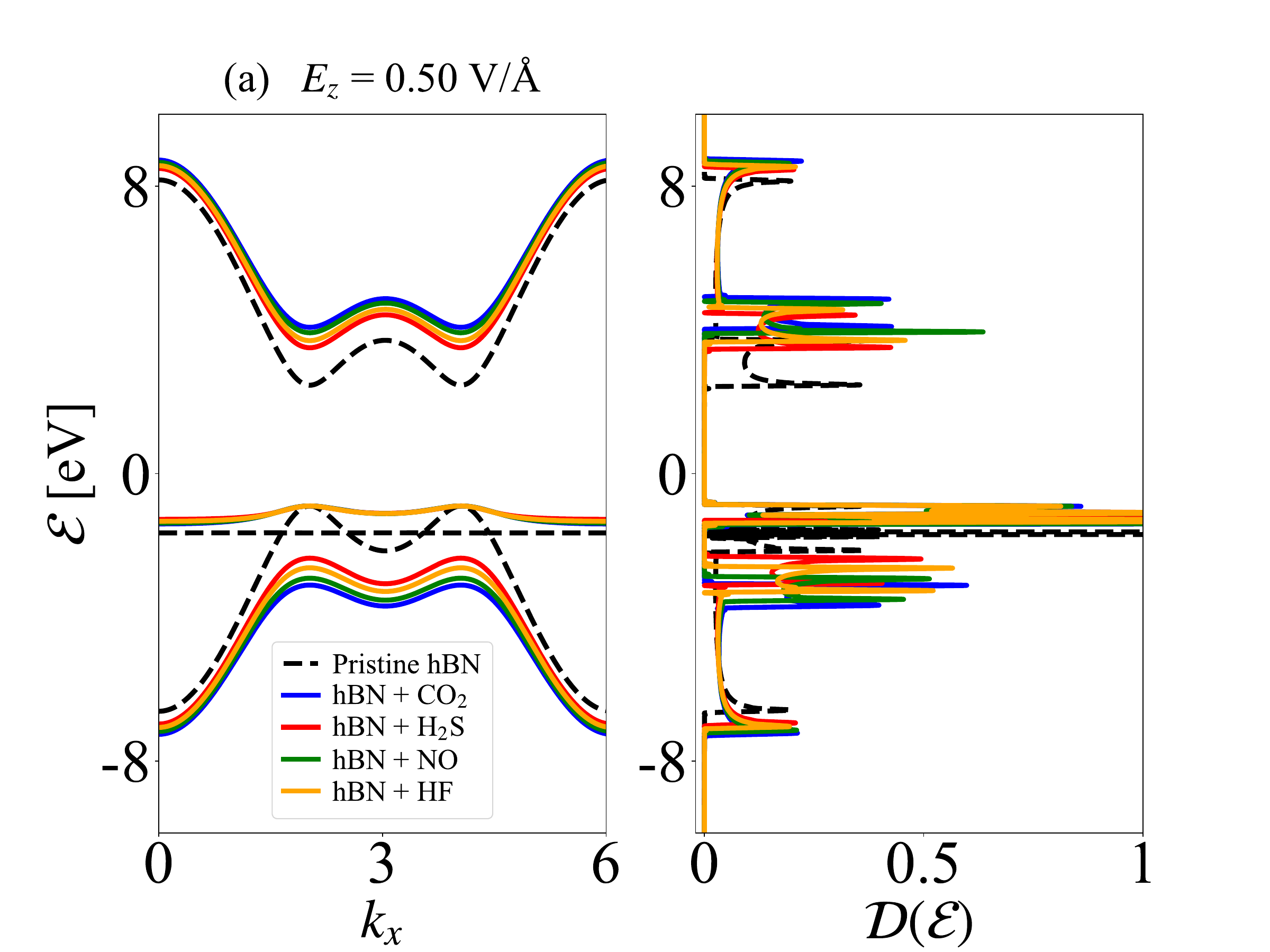}
\hspace{-11.3mm}
\includegraphics[scale=0.23]{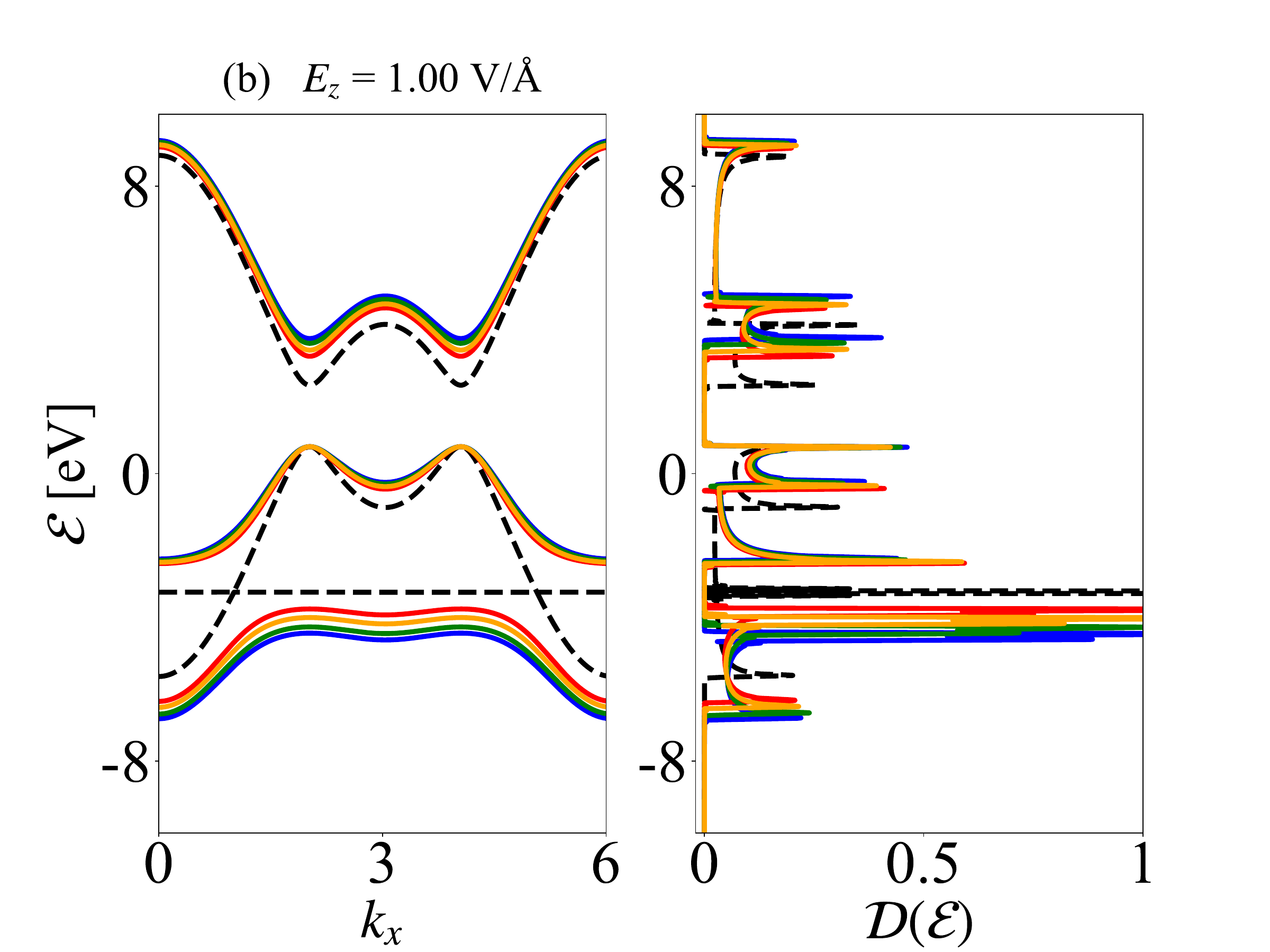}
\caption{Band structure and DOS of h-BN based FET under electric fields: (a) $E_z = 0.50$ V/\AA, (b) $E_z = 1.00$ V/\AA.}
\label{fig3}
\end{figure*}

\subsection {DOS in absence of external biasing}
The electronic band structure and the DOS for pristine h-BN and in the presence of different adsorbed gas molecules in the absence of any external bias are presented in Fig. \ref{fig2}. As shown in Fig. \ref{fig2}(a), the band structure of pristine h-BN has a wide intrinsic band gap with symmetric dispersion of the valence and conduction bands around the Fermi level and no mid-gap states, consistent with its insulating nature and in agreement with previous reports \cite{museur,verma,watanabe,cassabois}. 
Upon gas adsorption, the band structure undergoes significant changes depending on the electronic nature and binding characteristics of the adsorbate. Notably, the mid-gap states are found to shift towards the valence band, effectively narrowing the band gap and creating potential conducting channels within the originally insulating h-BN system. It is observed that the adsorption of CO$_2$ gas cause the most severe alteration of the valence and conduction band edges followed by NO. As a result, drastic changes in the electronic structure of h-BN are observed. This result suggests a more severe interaction than that typically encountered in the physisorption process. Moreover, it points towards the probability of partial charge transfer or polarization-induced coupling, which may not involve covalent bonding but still leads to a large electrostatic perturbation. The adsorption of HF leads to a moderate distortion of the band structure and DOS near the band edges compared to pristine h-BN. This is attributed to the formation of localized impurity states arising from orbital hybridization and weak chemisorption, particularly involving the p-orbitals of nitrogen and boron atoms. The introduction of H$_2$S has a very minor change in the band structure, indicating a significantly weak physisorption effect. 

The corresponding DOS of the system is depicted in Fig. \ref{fig2}(b). The appearance of the Van Hove Singularities (VHS) at the center and symmetrically in upper and lower bands for pristine h-BN are consistent with symmetric large energy gap with no mid gap states as observed from the band structure in Fig. \ref{fig2}(a). Upon gas adsorption a significant change in the DOS are observed. Mid-gap states appear for all gas molecules studied, accompanied by the emergence of asymmetric VHS features. These changes reflect the distortion of the pristine band structure and indicate the formation of localized states and broken particle-hole symmetry due to adsorbate-induced perturbations. Such features are most prominent in the case of CO$_2$ followed by NO as they induce a dramatic asymmetry in the appearance of the VHS in the DOS. This indicates a redistribution of electronic density and changes in the local potential. Such band-edge and DOS perturbations are crucial in controlling the carrier density and conductivity. These results demonstrate that CO$_2$ induces the strongest perturbation followed by NO while HF and H$_2$S induces moderate and weakest response which is in accordance with the previous works \cite{phung,kim2013,kim2024}. The gas-specific changes provide further evidence for the potential of h-BN as a selective and tunable gas sensor.

\subsection{Tuning the DOS via bias voltage}
To incorporate the effect of an out-of-plane electric field in the monolayer h-BN system, we consider a gate-induced potential difference $V_\text{gs}$ applied across the atomic thickness of the channel. Although the system consists of a single h-BN layer, the vertical field induces a linear Stark shift in the onsite energies of orbitals located at different positions along the $z$-axis, such as those associated with boron, nitrogen and possible gas-modified hybrid states. This leads to an effective breaking of mirror symmetry and modifies the electronic structure of the h-BN. To capture these effects, we modify the Hamiltonian Eq. (\ref{eq10}), to the form:
\begin{equation}
\mathcal{H}(\vec{k}) =
\begin{bmatrix}
E_0 + E_z & tf(\vec{k}) & 0 \\
tf^*(\vec{k}) & E_1 & t' \\
0 & t' & E_0' - E_z
\end{bmatrix}
\label{eq24}
\end{equation}
where, $E_z = V_\text{gs}/d$ is the resulting vertical electric field with $V_\text{gs}$ is the gate-source voltage and $d$ is the effective thickness of the h-BN monolayer. The field induces a linear electrostatic potential across the atomic orbitals, leading to a Stark shift in onsite energies of magnitude $\pm eV_\text{gs}/2$. This approach provides a tractable yet physically accurate representation of gate-tunable band alignment and carrier redistribution within the monolayer. Moreover, this framework captures the essential effects of vertical electric fields on the orbital degrees of freedom without invoking a multilayer structure and is consistent with the atomically thin geometry of the device and its pronounced field sensitivity. Fig. \ref{fig3}  illustrates the electronic band structure and the DOS of h-BN-based FET devices under applied vertical electric fields, in both pristine and gas-adsorbed configurations. Plots in Fig. \ref{fig3}(a) and (b) are respectively  for electric field strengths of $ E_z = 0.50  \text{V/\AA} $  and $ E_z = 1.00  \text{V/\AA} $.  As observed from Fig. \ref{fig2}, h-BN exhibits a large band gap with no mid-gap states in the absence of an electric field; however, the application of a vertical electric field induces a Stark effect, manifesting as a linear shift in onsite energies across the monolayer due to the potential difference across the monolayer causes a linear shift in onsite energies proportional to the atomic $z$-coordinate. As a result we have band gap renormalization, symmetry breaking and band edge asymmetry for $ E_z = 0.50 \text{V/\AA} $ as observed from Fig. \ref{fig3}(a). Additionally, band gap narrowing and asymmetry between the conduction and valence bands are observed. This is a consequence of the field-induced electrostatic potential that breaks the mirror symmetry along the h-BN plane and lifts the degeneracy between top and bottom layers. The effect is more pronounced at $ E_z = 1.00  \text{V/\AA} $ in Fig. \ref{fig3}(b), where the band gap is significantly reduced and both valence and conduction bands are more dispersive, indicating enhanced field-induced carrier mobility \cite{sharma}.

When gas molecules are adsorbed, the combined effect of electric field and gas-induced perturbation further modulates the band structure. CO$_2$ continues to produce the largest band shifts followed by NO, consistent with their polar character and strong coupling to the h-BN surface potential. Under finite $E_z$, this results in enhanced band bending, making them even more electronically active. In contrast, HF induces a moderate band shift that persists even under increasing electric field strength, indicating the robustness of localized hybridized states arising from orbital overlap between the gas molecules and the h-BN system. These states become more pronounced in the DOS, particularly at $E_z = 1.00,\text{V/\AA}$, where they begin to accumulate near the band edges, thereby enhancing carrier injection pathways. These results demonstrate that the interplay between vertical electric fields and gas adsorption provides a powerful mechanism for modulating the electronic response of h-BN monolayers, supporting their application as tunable and high-resolution chemical sensors.

\begin{figure*}[t]
\centerline
\centerline{ 
\includegraphics[scale=0.4]{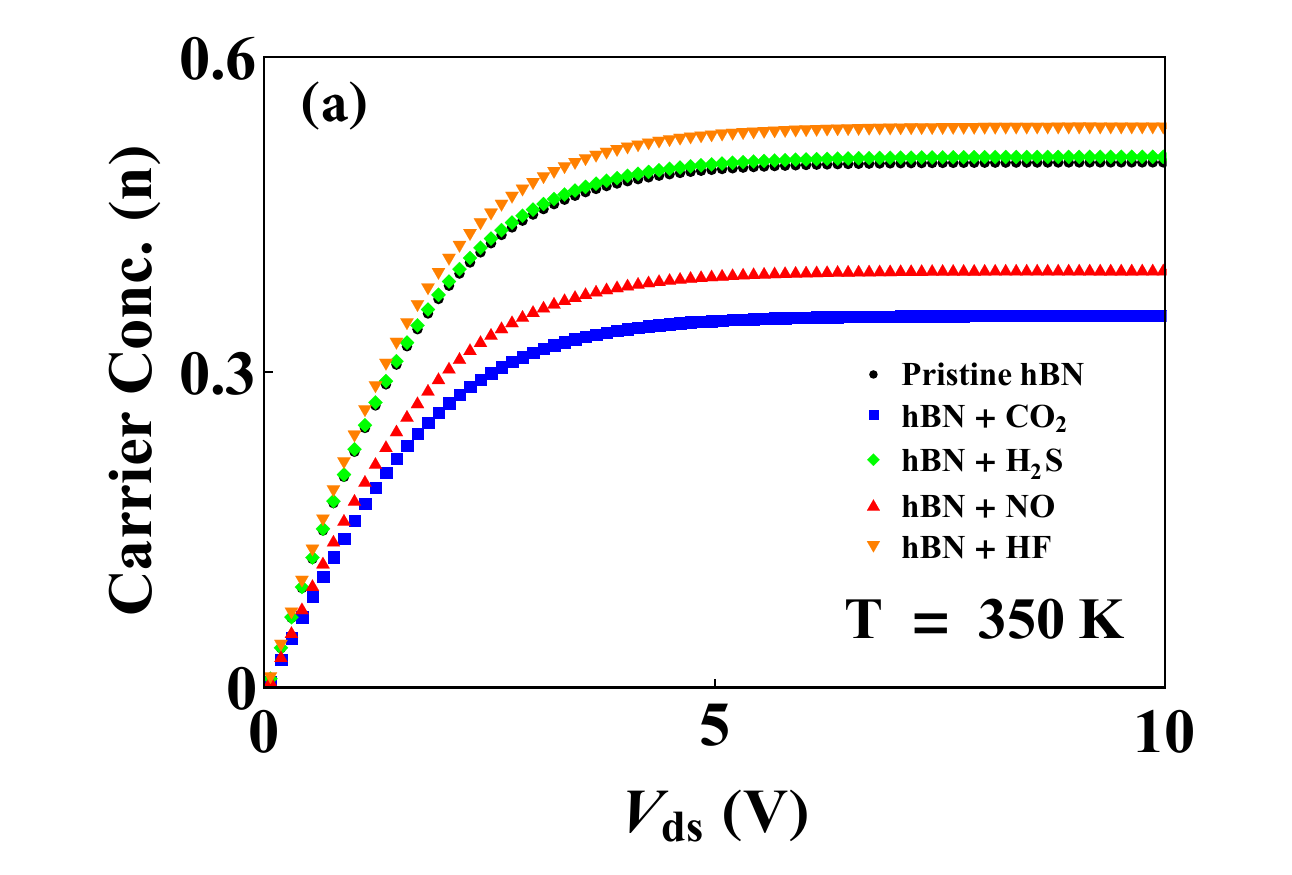}
\hspace{-10mm}
\includegraphics[scale=0.4]{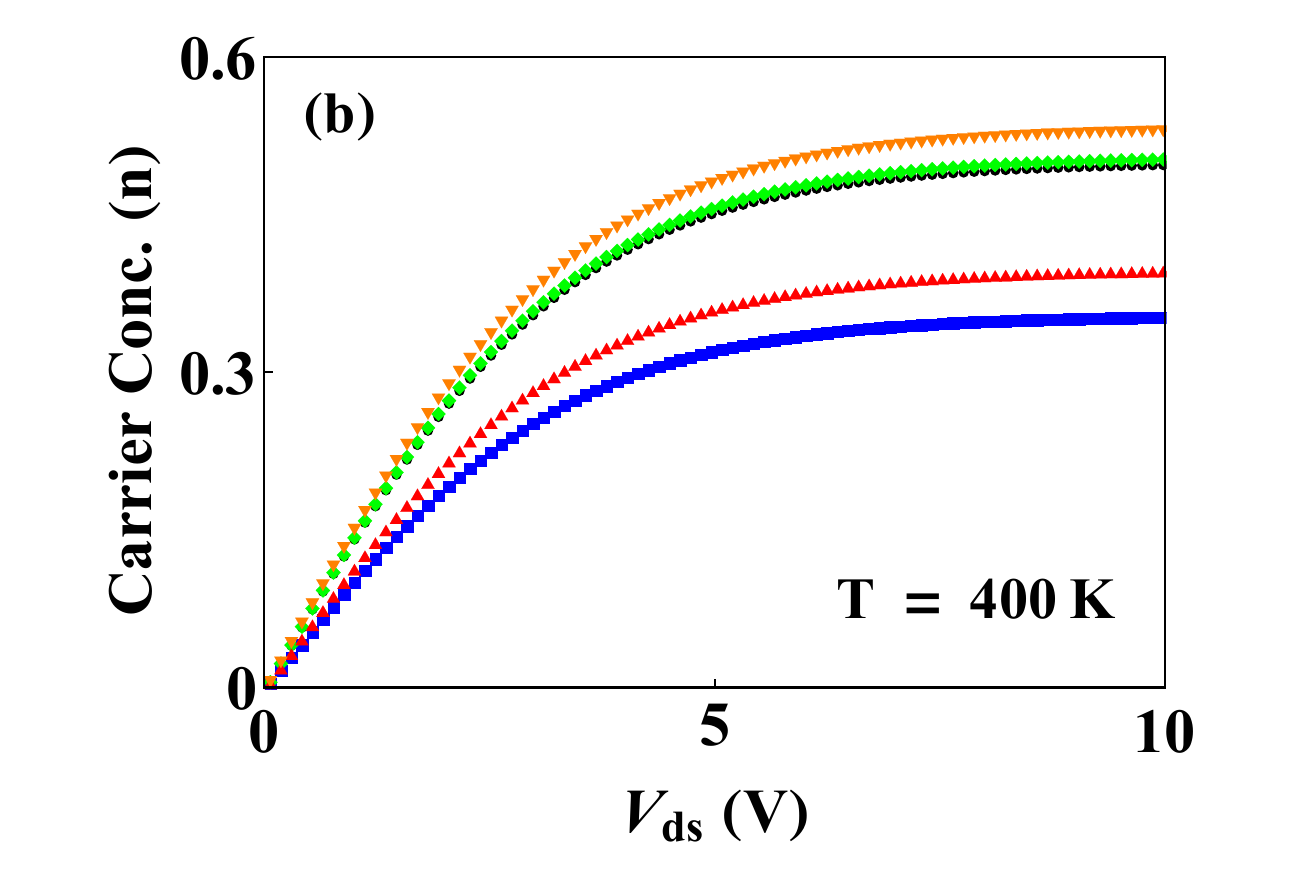}
}
\centerline{
\includegraphics[scale=0.36]{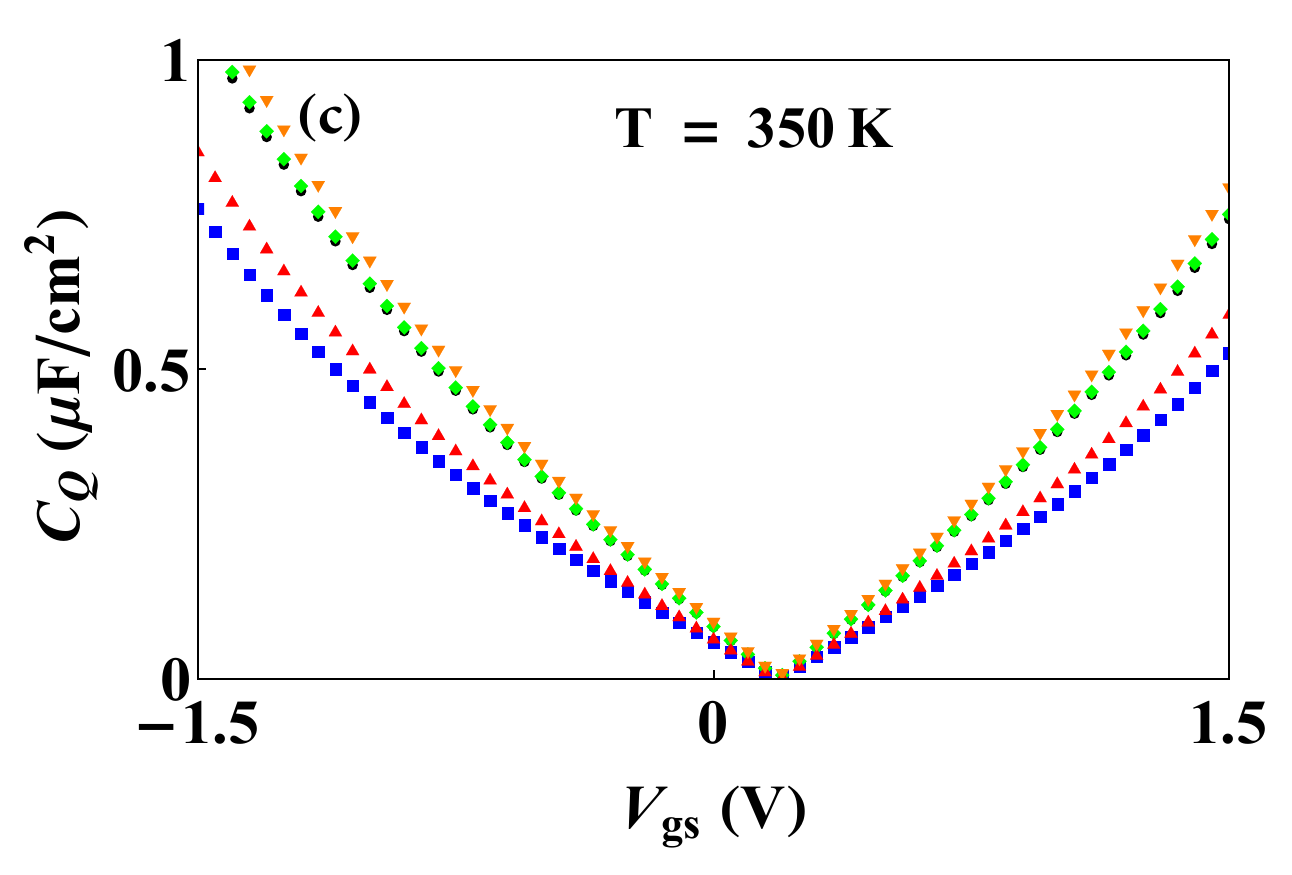}
\hspace{-3mm}
\includegraphics[scale=0.36]{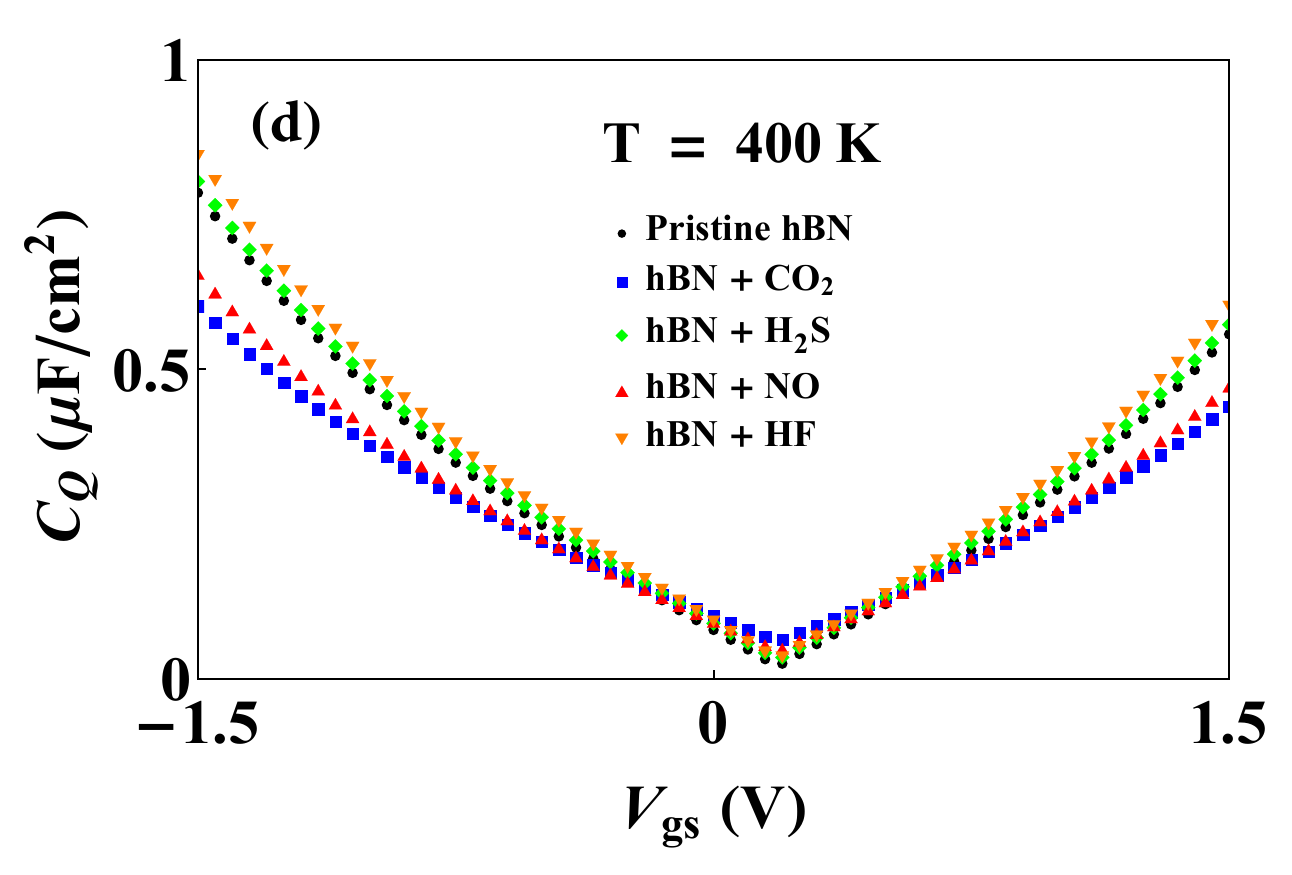}}
\caption{(Top Panel) Variation of carrier concentration with $V_{ds}$ of h-BN in absence and in presence for different adsorbed gases at (a) $350K$ (b) $400K$. (Bottom Panel) Quantum
capacitance of hBN as a function of $V_{gs}$ in absence and in presence of adsorbed gas molecules at (c) $350K$ and (d) $400K$.}
\label{fig4}
\end{figure*}
\section{Carrier concentration and Quantum Capacitance}

Adsorption of a gas molecule on the surface of h-BN can effectively alter the carrier concentration in the transistor channel. This result a change in quantum capacitance between the h-BN layer and the gate electrode. Thus, quantum capacitance serves as a critical parameter in studying the electrical response of a gas sensor and detecting gas molecules. It can be defined as \cite{dutta2005,acharjee2}:
\begin{equation}
    C_Q = \frac{\partial Q}{\partial V_\text{gs}}
    \label{eq25}
\end{equation}
where $Q$ is the total channel charge and $V$ is the applied gate voltage. This can be rewritten using the relation:
\begin{equation}
    C_Q = e^2 \frac{\partial n}{\partial \mathcal{E}}
    \label{eq26}
\end{equation}
where $e$ is the elementary charge and $n$ is the carrier concentration.

The carrier concentration $n$ of the system is given by \cite{dutta2005}:
\begin{equation}
    n = \int \mathcal{D}(\mathcal{E}) f(\mathcal{E}) \, d\mathcal{E}
    \label{eq27}
\end{equation}
where $f(\mathcal{E})$ is the Fermi-Dirac distribution function:
\begin{equation}
    f(\mathcal{E}) = \frac{1}{1 + \exp\left( \frac{\mathcal{E} - \mathcal{E}_\text{F}}{k_B T} \right)}
    \label{eq28}
\end{equation}
Here, $\mathcal{E}_\text{F}$ is the Fermi energy, $k_B$ is Boltzmann constant and $T$ is the temperature.

\begin{figure*}[t]
\centerline
\centerline{ 
\hspace{-0.5mm}
\includegraphics[scale=0.35]{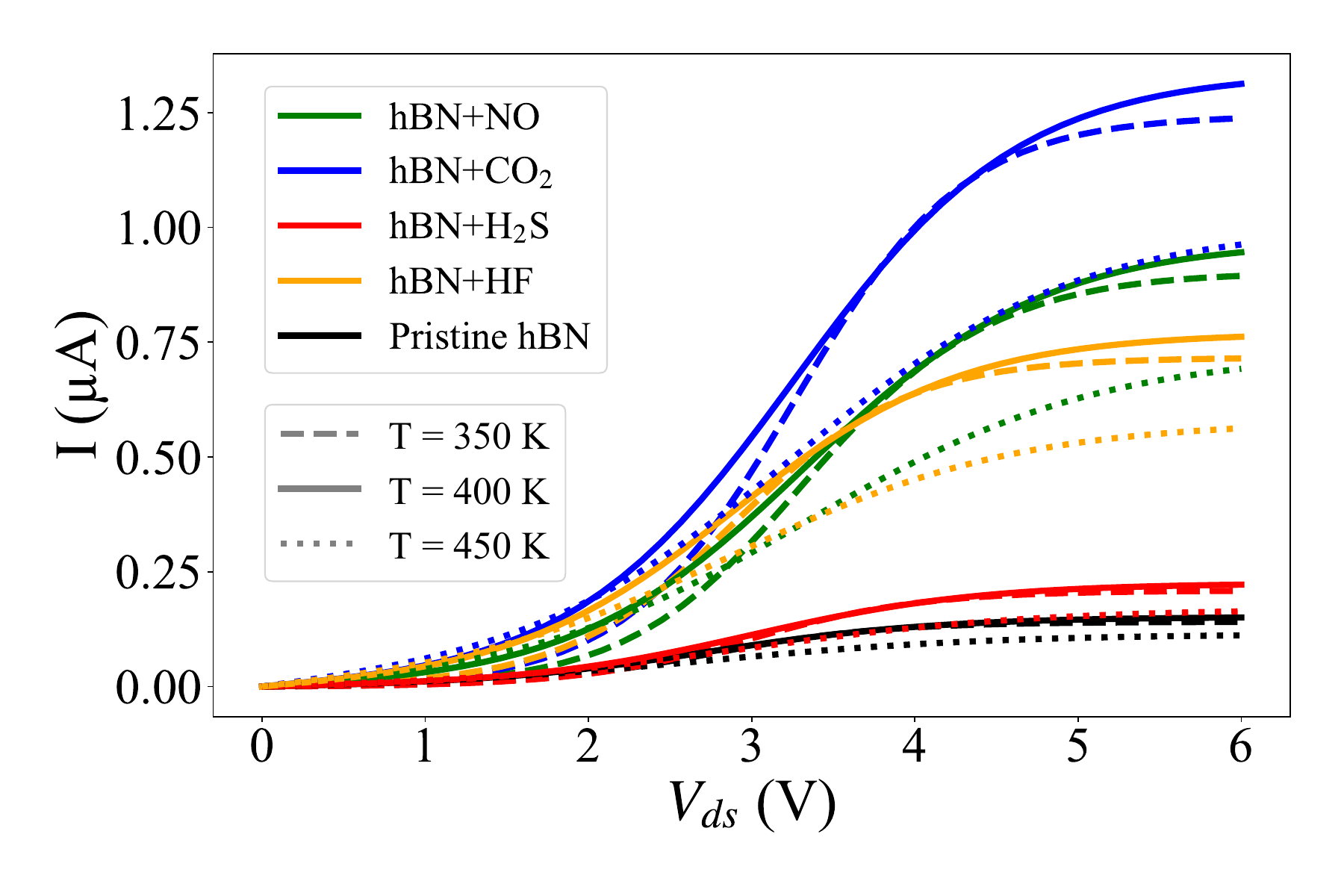}
}
\caption{I$-$V characteristics of h-BN  based FET in absence and in presence for different gases at different temperatures.}
\label{fig5}
\end{figure*}
To understand the impact of temperature and gas adsorption on carrier transport in h-BN based field-effect transistors, we plot the carrier concentration as a function of drain-source voltage ($V_{\mathrm{ds}}$) at $T = 350 K$ and $400 K$ in Fig. \ref{fig4}. The study compares for pristine h-BN with samples exposed to various gas molecules (CO$_2$, H$_2$S, NO, HF), each of which interacts differently with the h-BN surface due to their distinct charge transfer characteristics and binding energetics. Fig. \ref{fig4}(a) illustrate that at $T = 350 K$, carrier concentration increases with $V_\text{ds}$ for all systems, with H$_2$S  and HF  have higher carrier densities compared to CO$_2$, NO, and pristine h-BN. This suggests that H$_2$S and HF act as electron donors, enhancing the electron density in the conduction band via physisorption-induced charge transfer. In contrast, CO$_2$ and NO display relatively lower carrier concentrations indicating their strong electron acceptor characteristics, but have drastic shift from the pristine h-BN characterizing its strong sensing potential. For $T = 400 K$ [Fig. \ref{fig4}(b)], two key characteristics are observed: (i) the overall carrier concentration decreases for all exposed gases and (ii) the onset of saturation shifts to higher $V_{\mathrm{ds}}$. This behavior reflects thermally activated desorption of adsorbed molecules, which follows an Arrhenius behavior, $R_{\mathrm{des}} \propto \exp(-E_b/k_B T)$, where $E_b$ is the binding energy. Consequently, fewer charge carriers are induced at higher temperatures, leading to both reduced carrier concentration and delayed saturation with $V_{\mathrm{ds}}$. As temperature increases, the gases like NO and CO$_2$ desorb more readily, reducing their doping effect and restoring the carrier concentration closer to the pristine state. Notably, the impact of donor gases HF and H$_2$S is still visible at $400 K$, albeit diminished, highlighting their relatively stronger binding and persistent charge transfer.

Figs. \ref{fig4}(c) and \ref{fig4}(d) in the lower panel present the variation of quantum capacitance $C_Q$ with $V_\text{gs}$, which serves as a sensitive probe for the local density of states near the Fermi level in pristine and gas-adsorbed h-BN. Although pristine h-BN possesses a symmetric band structure, the charge neutrality point does not found to coincide with $V_{\mathrm{gs}} = 0$ due to the effect of extrinsic factors such as contact-induced doping, substrate effects and built-in electrostatic fields commonly present in realistic device environments.  At $T = 350 K$, adsorption of H$_2$S and HF results in narrower $C_Q$ profiles compared to the pristine case. This is consistent with enhanced carrier screening and increased DOS near the conduction band edge. Conversely, adsorption of CO$_2$ and NO leads to broader $C_Q$. With a rise in temperature to $400 K$ [Fig.~\ref{fig4}(d)], all $C_Q$ curves are greatly broadened and flattened, consistent with a reduction in surface-adsorbed gas molecules due to thermally activated desorption. This desorption hinders the doping effect, creating a more intrinsic response dominated by the wide bandgap nature of pristine h-BN. The temperature-induced broadening and flattening of $C_Q$ profiles therefore indicate a loss of gas selectivity and reduced DOS modulation, highlighting the necessity of optimized sensing temperature to maintain doping contrast and sensitivity.

\section{I-V Characteristics}

To study the charge transport properties of the hBN-based FET under finite bias, we employ the NEGF formalism. The applied drain-source voltage $ V_{ds} $ is incorporated by the shift of source (left) and drain (right) lead chemical potentials as $ \mu_L = \mathcal{E}_F + eV_{ds}/2 $ and $ \mu_R = \mathcal{E}_F - eV_{ds}/2$, where $ \mathcal{E}_F $  is the equilibrium Fermi level. The applied voltage bias opens a new window between the leads through which electrons can be transmitted resulting current flow through the device.

The current flowing through the system can be determined by using the Landauer–B\"{u}ttiker formalism, given by the Caroli formula \cite{caroli}
\begin{equation}
I(V_{ds}) = \frac{2e}{h} \int_{-\infty}^{\infty} T(\mathcal{E}, V_{ds}) [f_L(\mathcal{E}) - f_R(\mathcal{E})] d\mathcal{E},
\label{eq29}
\end{equation}
where $ f_{L,R}(\mathcal{E}) = \left[1 + \exp\left(\frac{\mathcal{E} - \mu_{L,R}}{k_B T}\right)\right]^{-1} $ are the Fermi-Dirac distribution functions for left and right electrode respectively and $ T(\mathcal{E}, V_{ds}) $ is the bias and energy dependent transmission function. The factor 2 arises to account the spin degeneracy.

The transmission function $ T(\mathcal{E}, V_{ds})$ is derived from the retarded Green's function of the device and the lead coupling can be written as \cite{keldysh}
\begin{equation}
T(\mathcal{E}, V_{ds}) = \mathrm{Tr} \left[ \Gamma_L(\mathcal{E}) \mathcal{G}(\mathcal{\mathcal{E}}) \Gamma_R(\mathcal{E}) \mathcal{G}^\dagger(\mathcal{\mathcal{E}}) \right]
\label{eq30}
\end{equation}
Here, the bias dependence of  $T(\mathcal{E}, V_{ds})$ arises from the potential drop across the device, which may be modeled as linear or spatially resolved depending on the device geometry and screening length. In the present analysis, we assume a symmetric potential drop and treat the transmission function as independent of voltage to leading order. 

Fig. \ref{fig5} illustrates the drain current ($I$) as a function of the drain-source voltage ($V_{ds}$) at different temperatures for pristine and h-BN in presence of adsorbed gases. In its pristine state, monolayer h-BN exhibits very low current due to its large band gap, which suppresses intrinsic carrier generation and results in a highly resistive channel. The I-V characteristics remains nonlinear, a hallmark of field-effect transistor (FET) operation owing to the gate modulation of carrier density and the non-Ohmic injection across the source-channel interface. Upon adsorption of gas molecules, the current increases significantly. It is observed that the exposure of CO$_2$ significantly change the I-V characteristics in comparison to pristine h-BN. A moderate change in I-V characteristics are observed for NO and HF while a very weak response is observed under adsorption of H$_2$S. 
With the increase in temperature the current also increases, indicating thermally activated transport. This behavior can be governed by the Boltzmann approximation for thermally excited carriers, where the conductance scales as $G \propto \exp\left(-\frac{E_a}{k_BT}\right)$, with $E_a$ representing the activation energy lowered by impurity states introduced via adsorption. The observed behavior is consistent with transport through localized states, where increased thermal energy enables carriers to hop or tunnel across gas-induced potential fluctuations.

\section{Sensitivity and Selectivity}
\subsection{Sensitivity}
The sensitivity $S$ of the device toward a specific gas is defined as the relative change in current upon adsorption \cite{acharjee2}.
\begin{equation}
    S = \frac{\Delta I}{I_0} = \frac{I_{\mathrm{gas}} - I_0}{I_0},
    \label{eq31}
\end{equation}
where, $I_{\text{gas}}$ is the current under the adsorption of gases while $I_0$ is the current for the pristine h-BN. As already discussed, the I-V characteristics in Fig. \ref{fig5} reveal significant changes upon adsorption of gases such as CO$_2$ and NO, indicating their potential sensitivity. 

\begin{figure}[t]
\centerline{ 
\includegraphics[scale=0.28]{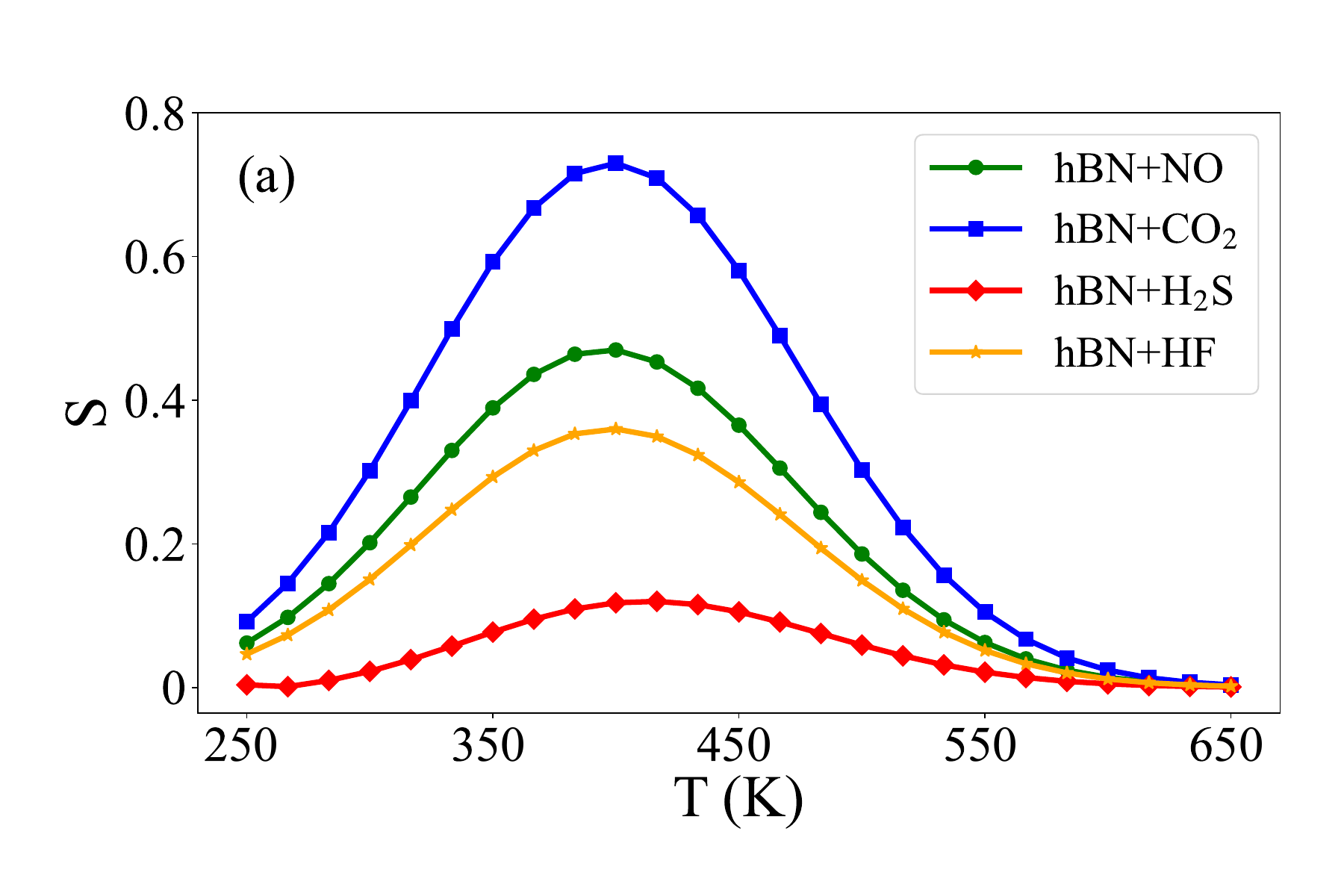}}
\centerline{ 
\includegraphics[scale=0.28]{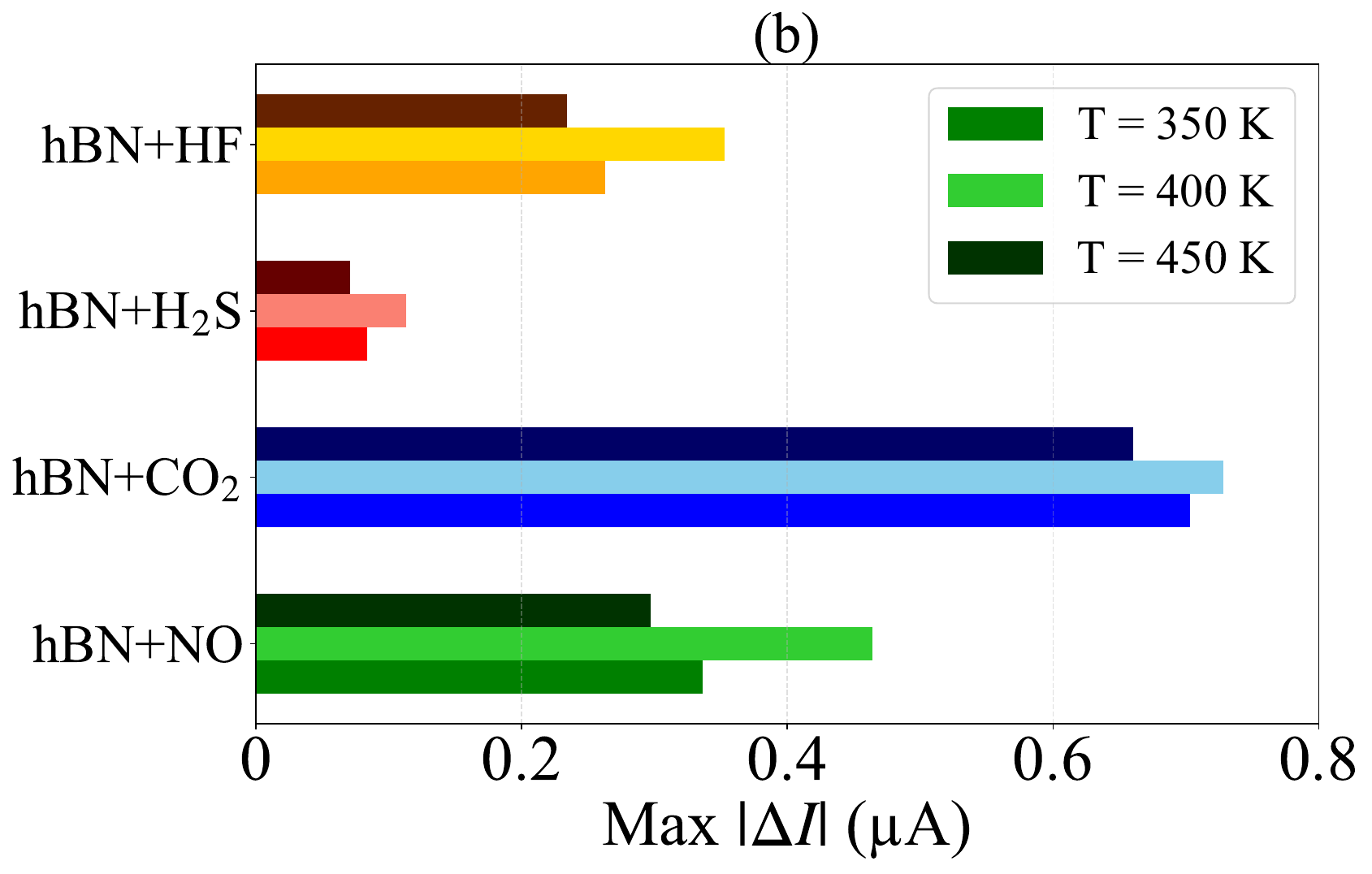}
}
\centerline{ 
\includegraphics[scale=0.28]{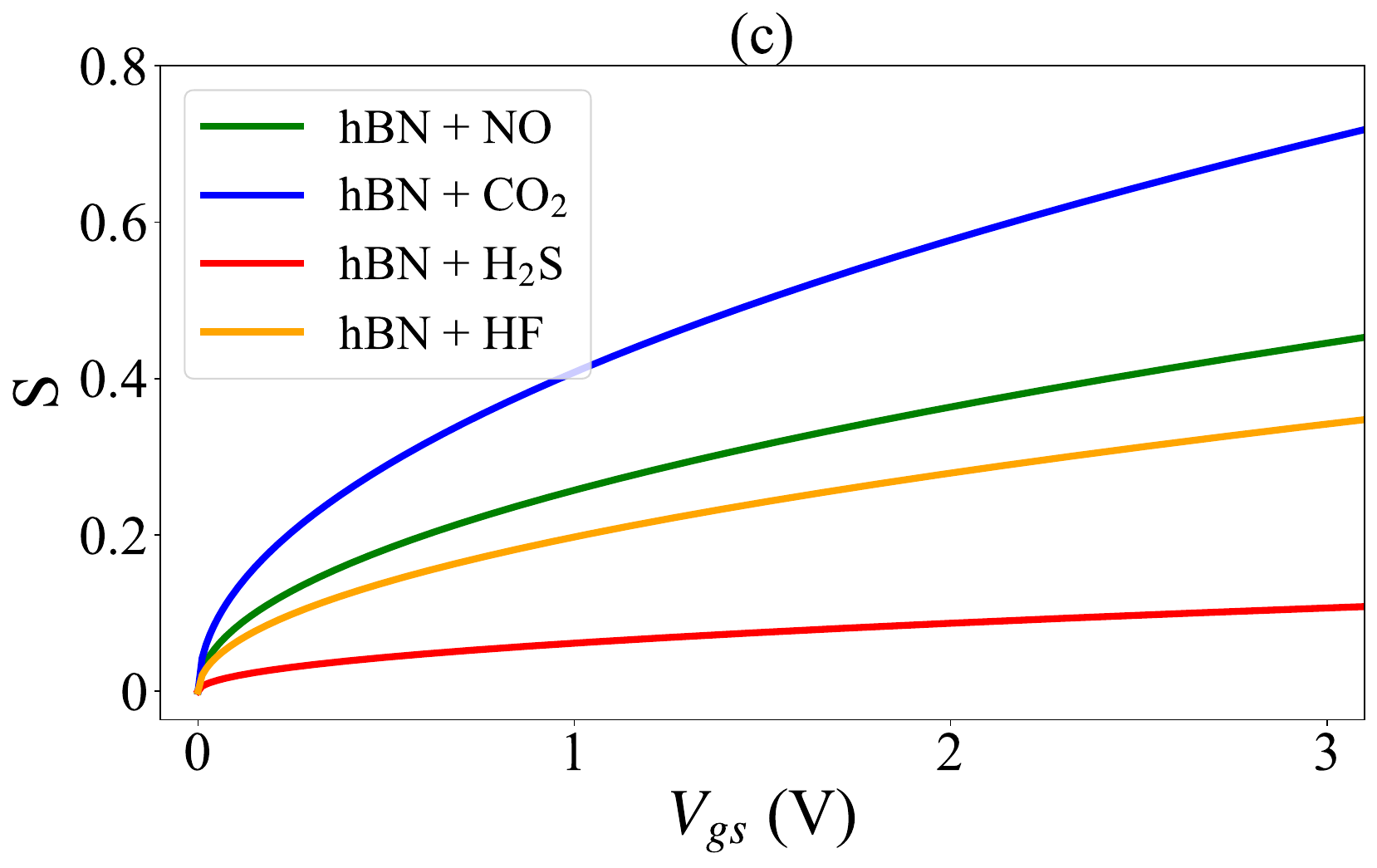}
}
\caption{(a) Temperature-dependent sensitivity profile of gas-adsorbed h-BN. (b) Maximum $|\Delta I|$ response for different gases at $300 K$, $350 K$, and $400 K$.  (c) Variation of sensitivity with $V_{gs}$ at $T = 400K$.}
\label{fig6}
\end{figure}

To further understand the influence of external conditions, we investigate the role of temperature and $V{\text{gs}}$ on sensitivity in Fig. \ref{fig6}. In Fig. \ref{fig6}(a), variation of sensitivity with temperature are is plotted as a function of temperature for different adsorbed gases. We observe a non-monotonic trend with a pronounced maximum around $\sim 410 K$, thereby a sudden fall is observed with further rise in temperature for all gases. This peak reflects a competition between two dominant physical processes: (i) enhanced charge transfer at elevated temperatures, due to increased thermal energy allowing electrons from gas-induced states to populate the h-BN conduction band and (ii) thermal desorption, which becomes dominant above $450 K$ and reduces the effective surface coverage of adsorbed molecules. Among the gases, CO$_2$ exhibits the highest sensitivity, followed by NO and HF, while H$_2$S display the weakest response. This ordering aligns with earlier observations from band structure modification, DOS and quantum capacitance trends shown in Fig.\ref{fig2} and Fig.\ref{fig4}, indicating a direct correlation between the electronic perturbation induced by the gas molecule and the overall sensor performance. 

Fig. \ref{fig6}(b) presents the maximum absolute current modulation, defined as $\Delta I = |I_{\text{gas}} - I_0|$, for each gas at three different temperatures: $350 K$, $400 K$ and $450 K$.  The current modulation $\Delta I$ is found to be maximized at $400 K$ for all gases,  consistent with the sensitivity peak in Fig. \ref{fig6}(a), It indicates the optimal charge transfer regime before thermal desorption effects dominate. The observed gas sensitivity are found to follow the order CO$_2$ $>$ NO $>$ HF $>$ H$_2$S, persists across all temperatures, confirming that the strength of interaction between the adsorbed molecules and the h-BN channel primarily governs the sensor response. For  $450 K$, a sharp decline in $\Delta I$ is observed, highlighting the thermal desorption that reduces effective surface coverage. Notably, CO$_2$ exhibits the highest modulation in drain current, especially at $400 K$, suggesting both a stronger adsorption energy and better energetic alignment between its molecular orbitals and the conduction band edge of h-BN.

The variation of sensitivity $S$ with $V_{\text{gs}}$ for different adsorbed gases at $T = 400 K$ is shown in Fig. \ref{fig6}(c). The application of $V_{\text{gs}}$ modulates the electrostatic potential in the h-BN channel, effectively shifting the Fermi level and thereby change the carrier concentration. With the increase in $V_{\text{gs}}$, additional conduction band states become accessible, thereby enhancing charge transfer from gas-induced donor or acceptor levels to the channel. This leads to a non-linear increase in sensitivity, particularly steep at low voltages due to field-induced band bending near the interface. For higher $V_{\text{gs}}$, sensitivity tends to saturate especially for weakly interacting gases like HF and H$_2$S.  This is due to limited additional modulation in surface potential and minimal change in the DOS. Notably,  CO$_2$ and NO exhibit the steepest increase and have highest absolute sensitivity, which is consistent with the results in Figs. \ref{fig6}(a) and \ref{fig6}(b). Thus the gate field  serves as an effective tuning parameter, amplifying gas-induced effects through modulation of band alignment and quantum capacitance.

\begin{figure*}[hbt]
\centerline{ 
\includegraphics[scale=0.3]{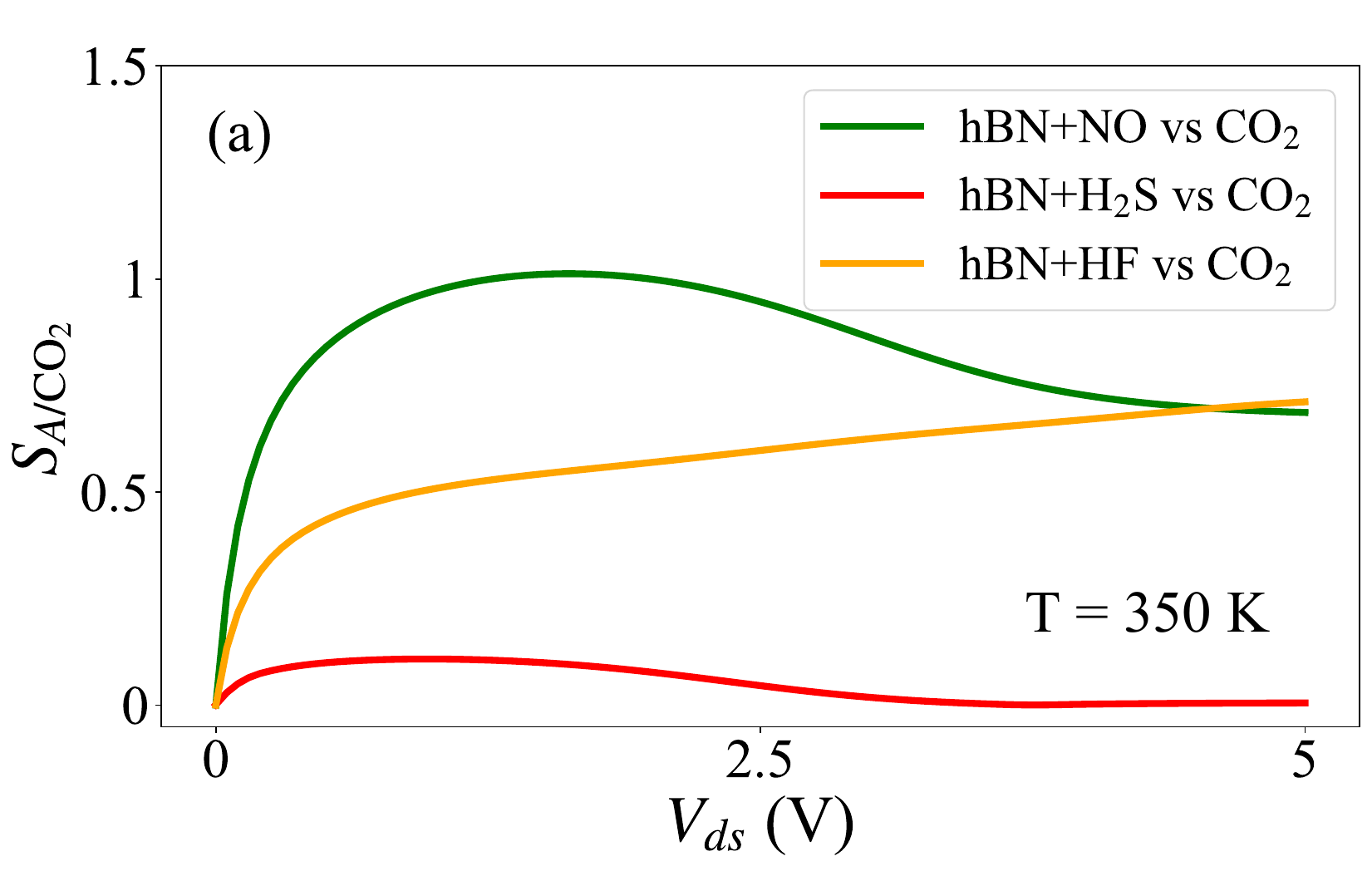}
\includegraphics[scale=0.3]{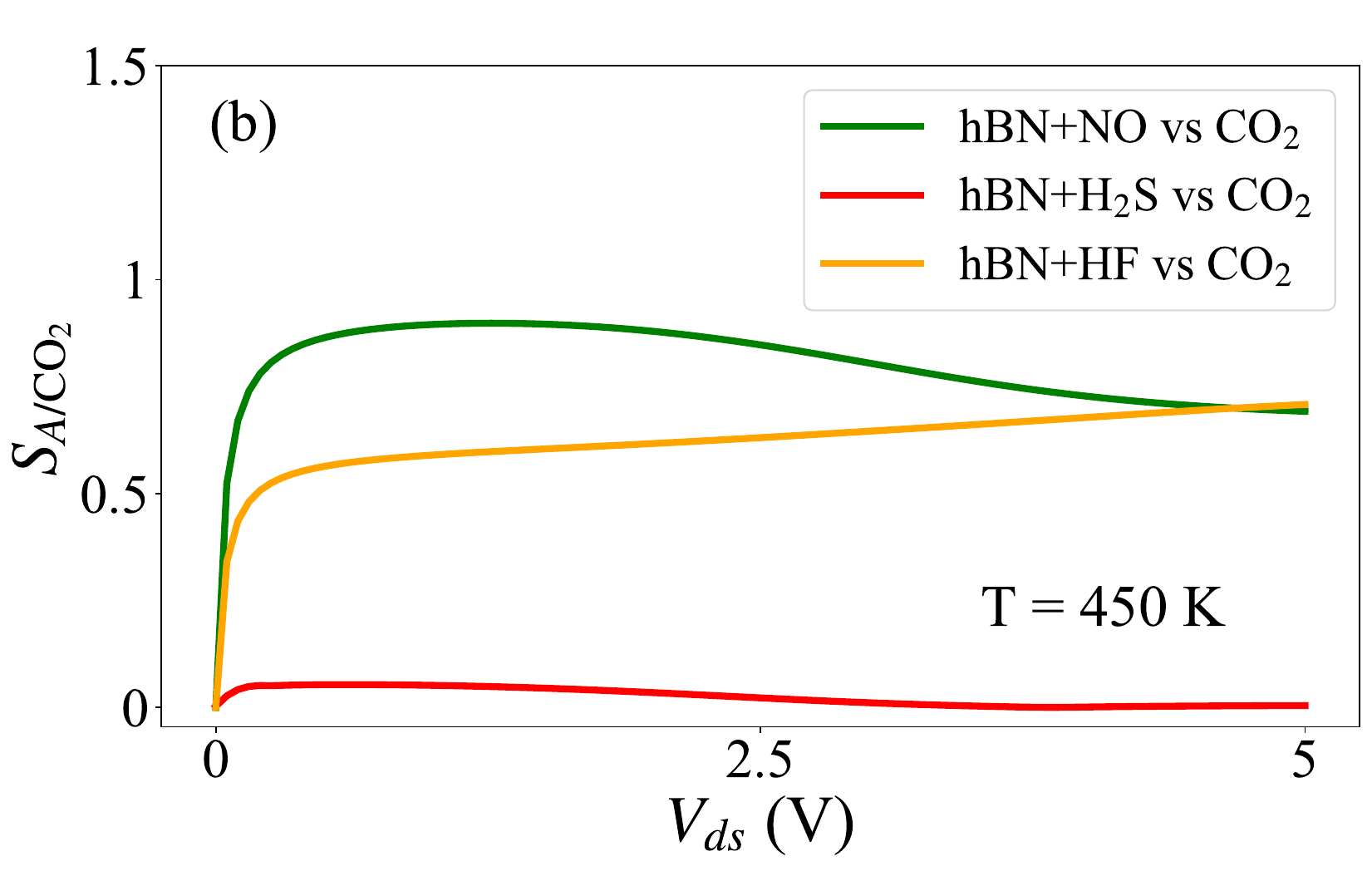}
}
\caption{(a) Variation of Selectivity $S_{A/\mathrm{CO_2}}$ with $V_{ds}$ for different gases considering CO$_2$ as reference at two different temperatures: (a) $300 K$ and (b) $400 K$.  }
\label{fig7}
\end{figure*}

\subsection{Selectivity}

The selectivity ratio  $S_{\text{gas}/\mathrm{CO_2}}$  quantifies the relative sensing response of a target gas compared to CO$_2$, serves as a key metric for determining the ability of our h-BN based FET device to distinguish between different gas species. It can be defined as
\begin{equation}
    S_{\text{X}/\text{CO}_2} = \frac{|\Delta I_\text{X}|}{|\Delta I_{\mathrm{CO_2}}|} = \frac{|I_{\mathrm{X}} - I_0|}{|I_{\mathrm{CO_2}} - I_0|}
\label{eq32}
\end{equation}
Fig. \ref{fig7} presents the variation of selectivity $S_{\text{X}/\text{CO}_2}$ with $V_\text{ds}$ for different gases (X = NO, H$_2$S, HF) relative to CO$_2$ at temperatures: $350 K$ and $450 K$. We consider CO$_2$ as our reference gas due to its consistently strong adsorption and highest sensitivity at different temperatures. We observe that at both temperatures, NO exhibits the highest relative selectivity over CO$_2$, particularly at low bias condition. However, for $T = 350 K$ at $V_\text{ds} \sim 2.2V$, the selectivity ratio of NO becomes 1 indicating non distinguishability of NO over CO$_2$ in moderate $V_\text{ds}$ condition. For $V_\text{ds} > 2.2V$, $S_{\text{NO}/\text{CO}_2}$ shows a non-monotonic decrease indicating NO has weaker response than CO$_2$. It is observed that 
$S_{\text{H}_2\text{S}/\text{CO}_2} < S_{\text{HF}/\text{CO}_2}$ for all $V_\text{ds}$, indicating HF has stronger response than H$_2$S. Although $S_{\text{NO}/\text{CO}_2} > S_{\text{HF}/\text{CO}_2}$ for low and moderate $V_\text{ds}$ but at higher $V_\text{ds}$, the two gases exhibit comparable responses at higher biases, indicating reduced selectivity under strong electric field as observed from Fig. \ref{fig7}(a). With the increase in temperature to $450 K$ a slight decrease in selectivity is observed for all gases. Moreover, $S_{\text{NO}/\text{CO}_2} < 1$ for all values of $V_\text{ds}$, indicating that CO$_2$ exhibits a stronger sensing response than NO at high temperatures. The slightly enhanced selectivity contrast at $450 K$ stems from increased thermal desorption of weakly bound species like H$_2$S, which further suppresses their influence on current modulation, thereby amplifying selectivity differentials. These results underscore the importance of both bias tuning and temperature control in achieving selective gas sensing using h-BN FET platforms.

Our quantum transport based findings using NEGF extend previous DFT based studies \cite{phung,kim2013,kim2024,rahimi,kalwar} by providing a complete device - level perspective on gas sensing in h-BN based FET. In contrast to prior work focused on static adsorption energetics, our model captures gate-tunable I-V characteristics, quantum capacitance variation, and temperature-dependent sensitivity, offering a realistic design platform for practical h-BN based gas sensors.

\section{Conclusion}
In summary, we have studied a comprehensive theoretical analysis of gas sensing in monolayer h-BN based FET using the non-equilibrium Green’s function formalism combined with the Landauer–B\"{u}ttiker transport approach. Our analysis goes beyond conventional DFT-based studies by incorporating a full device-level perspective, encompassing electronic structure modification, charge transport with field and temperature dependent sensitivity. By modeling, we investigate the effect of adsorbed gas molecules like CO$_2$, NO, HF, and H$_2$S on the band structure and DOS  of the monolayer h-BN. We observed that gas adsorption strongly perturbs the electronic band structure and DOS of h-BN, with CO$_2$ and NO inducing the most significant modifications through mid-gap state formation and band edge shifts. Moreover, these perturbations can enhance carrier modulation and hence the sensing response. Apart from that an application of a vertical electric field leads to band gap narrowing via the Stark effect, enhancing carrier mobility and allowing tunability of the gas response. When combined with gas adsorption, this field-induced modulation produces unique energy-dependent characteristics that are essential for gas distinction and selectivity. We found that  gas sensing characteristics are strongly dependent on temperature. At moderate temperatures, enhanced charge transfer from adsorbed molecules improves conductivity, while at higher temperatures thermal desorption dominates leading to decrease in carrier concentration and sensitivity. This temperature-dependent behavior, reflected in the variations of the I$-$V characteristics and quantum capacitance, reveals that CO$_2$ consistently exhibits the highest sensitivity, followed by NO and HF, whereas H$_2$S shows the weakest response. Additionally, we evaluated the selectivity of the h-BN based FET device by analyzing the relative sensitivity ratio for different gases considering CO$_2$ as a reference due to its dominant sensing signature. Our results highlight that CO$_2$ exhibits higher selectivity than NO at low bias and moderate temperature condition, however a non selective characteristics of NO with CO$_2$ is observed at moderate biasing condition. At higher biasing condition CO$_2$ regains dominance. Moreover, at higher temperatures, CO$_2$ remains highly selective for all gases while gases like HF and H$_2$S display limited selectivity and reduced modulation in transport characteristics. Furthermore, it is observed HF consistently outperforms H$_2$S under all conditions.  Overall, our work emphasizes the crucial role that band structure engineering, electrostatics, and thermodynamics play in maximizing sensor response and offers fresh perspectives on the design of tunable, field effect controlled gas sensors based on h-BN based channels.

\end{document}